\documentclass[11pt]{article}
\usepackage{moriond,epsfig}

\def\be{\begin{equation}}
\def\ee{\end{equation}}
\def\bea{\begin{eqnarray}}
\def\eea{\end{eqnarray}}

\begin{document}
\vspace*{4cm}
\title{GALAXY-GALAXY LENSING AS A PROBE OF \\ GALAXY DARK MATTER HALOS}

\author{M. LIMOUSIN$^1$, J.-P. KNEIB$^2$ \& P. NATARAJAN$^3$}

\address{$^1$ Dark Cosmology Centre, Niels Bohr Institute, Juliane Maries Vej 30, 2100 Copenhagen, Denmark\\
$^2$ OAMP, Laboratoire d'Astrophysique de Marseille, Traverse du siphon, 13012 Marseille, France\\
$^3$ Department of Astronomy, Yale University, 260 Whitney Avenue, New Haven, CT 06511, USA}

\maketitle\abstracts{
Gravitational lensing has now become a popular tool to measure the mass distribution 
of structures in the Universe on various scales.
Here we focus on the study of galaxy's scale dark matter halos with galaxy-galaxy lensing
techniques: observing the shapes of distant background galaxies which have been lensed by 
foreground galaxies allows us to map the mass distribution of the foreground galaxies.
The lensing effect is small compared to the intrinsic ellipticity distribution of galaxies,
thus a statistical approach is needed to derive some constraints on an average lens population.
An advantage of this method is that it provides a probe of
the gravitational potential of the halos of galaxies out to very large radii, where 
few classical methods are viable, since dynamical and hydrodynamical tracers of the 
potential cannot be found at this radii.
We will begin by reviewing the detections of galaxy-galaxy lensing obtained so far.
Next we will present a maximum likelihood analysis of simulated data we performed to evaluate
the accuracy and robustness of constraints that can be obtained on galaxy halo properties.
Then we will apply this method to study the properties of galaxies which stand in massive cluster 
lenses at $z\sim0.2$. The main result of this work is to find dark matter halos of cluster 
galaxies to be significantly more compact compared to dark matter halos around field galaxies of 
equivalent luminosity, in agreement with early galaxy-galaxy lensing studies and with theoretical
expectations, in particular with the tidal stripping scenario. We thus
provide a strong confirmation of tidal truncation from a homogeneous
sample of galaxy clusters. Moreover, it is the first time that cluster
galaxies are probed successfully using galaxy-galaxy lensing techniques
from ground based data.
}

This contribution is a summary of two galaxy-galaxy lensing papers:\\ Limousin et~al., 2005
\cite{paperI}, hereafter Paper I exposes a theoretical analysis of galaxy-galaxy lensing, 
and Limousin et~al., 2006 \cite{paperII}, hereafter Paper II exposes the application of the 
method tested extensively in Paper I on a sample of homogeneous massive galaxy cluster
at $z\sim0.2$.

\section{Galaxy-Galaxy Lensing}
\begin{figure}[h!]
\begin{center}
\includegraphics[height=7cm,width=7cm]{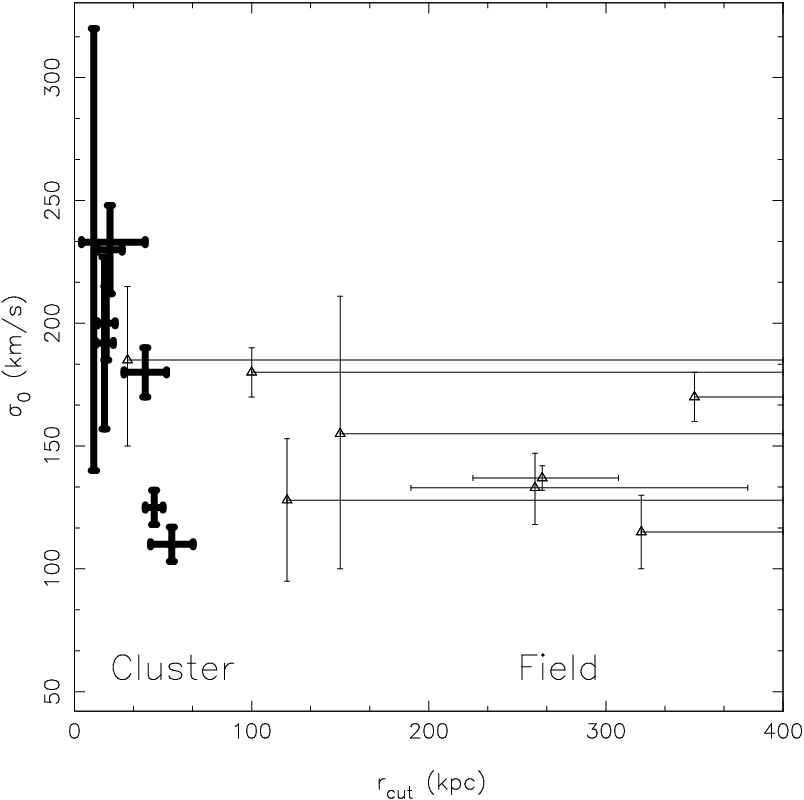}
\caption{Comparison between galaxy-galaxy lensing results on cluster galaxies
(black) and galaxy-galaxy lensing results on field galaxies (grey): cluster galaxies 
appear to be significantly more compact than field galaxies}
\label{comparfieldcluster}
\end{center}
\end{figure}

The main goal of galaxy-galaxy lensing studies is to obtain
constraints on the physical parameters that characterise the dark
matter halos of galaxies. A dark matter halo can be described by two
parameters: in this work we will mainly use $\sigma_0$, the central velocity dispersion,
which is related to the depth of the potential well, and
$r_{\mathrm{cut}}$, the cut off radius, which is related to the spatial
extension of the halo since it defines a change in the slope of the
three dimensional mass density profile: below $r_{\mathrm{cut}}$, the
profile falls with radius (see Paper I for a detailed description of
galaxy dark matter halo modeling). \\
%The first \emph{non} detection of galaxy-galaxy lensing is the work
%by \citet{tyson}. Despite a vast amount of data (about 28\,000
%foreground-background pairs), they were unable to get a galaxy-galaxy
%lensing signal, mainly because of the poor quality of the data at that time.
The first statistically significant detection of galaxy-galaxy lensing
is the work by Brainerd, Blandford \& Smail, 1996 \cite{bbs}, hereafter \textsc{bbs}.
They used deep ground-based imaging data
($\simeq$ 72 sq. arcminute) to investigate the orientation of 511 faint
background galaxies relative to 439 brighter foreground field galaxies.
They claimed a detection of galaxy-galaxy lensing on angular scales
between $5"$ and $35"$ and derived limits on the characteristic
parameters of the dark matter halos of L* field galaxies :
$\sigma_0=155 \pm 56$ km\,s$^{-1}$ and $r_{\mathrm{cut}} > 100h^{-1}$ kpc.
Since \textsc{bbs}, there have been 12 independent detections of galaxy-galaxy
lensing by field galaxies: 
Griffiths et~al., 1996 \cite{griffiths}; Dell'Antonio \& Tyson, 1996 \cite{dellantonio};
Hudson et~al., 1998 \cite{hudson98}; Ebbels et~al., 1998 \cite{ebbels};
Fisher et~al., 2000 \cite{fisher}; Jansen, 2000 \cite{jaunsen}; McKay et~al., 2001 \cite{mckay};
Smith et~al., 2001 \cite{smith01}; Wilson et~al., 2001 \cite{wilson}; 
Hoekstra et~al., 2002 \cite{hoekstracnoc2}; Kleinheinrich, 2003 \cite{martinacombo17};
Hoekstra et~al., 2004 \cite{hoekstra04}.
Galaxy-galaxy lensing has also been used successfully
to map substructure in massive lensing clusters 
(Natarajan et~al., 1998 \cite{priyaAC114}, 2002a \cite{priyaA2218}, 2002b \cite{priyatidalstrip};
Geiger \& Schneider, 1998 \cite{geigeramas}).
Analyses on cluster galaxies all used HST data for their investigations, and most of them
included strong constraints from observation of multiple images systems.\\
Fig.~\ref{comparfieldcluster} illustrates the results of the different detections in term
of constraints in the ($\sigma_0,r_{\mathrm{cut}}$) plane.
Before comparing these results, one has to keep in mind that there is considerable variation
between data sets and the analysis techniques used by the
various authors. The imaging quality, size of the field, the dichotomy 
between lenses and sources differs significantly amongst these
investigations. Moreover, the data are a heterogeneous mix of
deep images which were acquired for purposes other than galaxy-galaxy
lensing studies.  Despite these differences, the implications of these studies for the
physical characteristics of the halos of field galaxies are all
broadly consistent with one another, and the results on field galaxies are in 
good agreement with studies based on satellite dynamics (see \emph{e.g.} Prada et~al., 2003 \cite{prada}). \\
From Fig.~\ref{comparfieldcluster}, a clear trend can be seen: dark
matter halos in cluster are significantly more compact compared to
halos around field galaxies of equivalent luminosity.  This is a
landmark observational result provided by galaxy-galaxy lensing
studies that was expected from theoretical considerations and
numerical simulations: when clustering, galaxies do
experience strong tidal stripping from the cluster potential, and they
loose part of their dark matter halo, feeding the global cluster dark
matter halo itself.

\noindent We can say that the last ten years since the first detection of
galaxy-galaxy lensing have been 'experimental' in the sense that these
early studies have demonstrated convincingly that galaxy-galaxy
lensing, though challenging to detect, is a viable technique by which
the dark matter distribution on scales of individual galaxies can be
investigated. Now that the technique has been proved, galaxy-galaxy
lensing appears as very promising to get interesting
statistical constraints on galaxy physics (halo parameter determination, mass measurements,
M/L ratio and evolution with redshift and environment; deviation from spherical symetry for
dark matter halos; morphological dependence of the halo potential; scaling of the 
total galaxy mass with luminosity; truncation of dark matter halos during the infall
of galaxies into cluster; nature of dark matter - see Paper II for more details).

\section{Simulating Galaxy-Galaxy Lensing}
We first introduce some mass models to describe galaxies, and expose the simulation
procedure and the methodology. Then we present some results. More details can be found
in Paper I.
\subsection{Modeling the mass distribution of galaxies}
Two different mass models are used to model galaxy dark matter halos:
(i) the two components Pseudo-Isothermal Elliptical Mass Distribution 
(\textsc{piemd}, Kassiola \& Kovner, 1993 \cite{kassiola}),
which is a more physically motivated mass profile than 
the isothermal sphere profile (\textsc{sis}) but sharing the same slope
at intermediate radius and (ii) the ``universal'' \textsc{nfw} profile 
(Navarro, Frenk \& White, 1996 \cite{nfw}). 

\subsubsection{PIEMD profile:}
The density distribution for this model is given by:

\begin{equation}
\label{rhoPIEMD}
\rho(r)=\frac{\rho_0}{(1+r^2/r_{\mathrm{core}}^2)(1+r^2/r_{\mathrm{cut}}^2)}
\end{equation}
with the core radius $r_{\mathrm{core}}$ of the order of $0.1"$, and a
truncation radius $r_{\mathrm{cut}}$. In the centre,
$\rho\sim\rho_0/(1+r^2/r_{\mathrm{core}}^2)$ which describes a core with
central density $\rho_0$. The transition region
($r_{\mathrm{core}}<r<r_{\mathrm{cut}}$)
is isothermal, with $\rho\sim r^{-2}$. In the outer parts, the
density falls off as $\rho\sim r^{-4}$, as is usually required for
models of elliptical galaxies.
This mass distribution is described by a central velocity dispersion $\sigma_0$
related to $\rho_0$.
It is easy to show that for a vanishing core radius, the surface mass
density profile obtained above becomes identical to the surface mass
profile used by \textsc{bbs} for modeling galaxy-galaxy lensing. Since many
authors are using the same mass profile in their
galaxy-galaxy lensing studies, it allows easy comparison of our
results.
It should be noted that a dark matter halo parametrised by $r_{\mathrm{cut}}$
still have a significant amount of mass below $r_{\mathrm{cut}}$: the mass
profile become steeper, but about half of the mass is contained below
$r_{\mathrm{cut}}$. Thus $r_{\mathrm{cut}}$ can be considered as a half mass radius.

\subsubsection{NFW profile:}

The \textsc{nfw} density profile provides a good fit to the halos
that form in N-body simulations of collisionless dark matter, over 9 orders of 
magnitudes in mass, from the scale of globular clusters to that of massive galaxy
clusters. The density distribution of the \textsc{nfw} profile is given by:
\begin{equation}
\label{rhoNFW} 
\rho(r) =  \frac{\rho_s}{(r/r_{s})(1+r/r_{s})^{2}}
\end{equation}
where $r_s$ is a characteristic radius that defines a change in the slope
of the density profile.
This mass distribution is described by a characteristic velocity dispersion $\sigma_s$
related to $\rho_s$.
It can be also parametrised in terms of $M_{200}$, which is the mass
contained in a radius $r_{200}$ where the criterion
$\overline{\rho}=200\rho_{\mathrm{crit}}$ holds, and
$c=r_{200}/r_s$, the concentration parameter.

\subsection{Simulating images}
We describe foreground lenses by a mass profile with known input parameters, scaled
as a function of luminosity. 
We put the individual lenses constituting a cluster at a redshift of
0.2, and model it as a superposition of a large-scale smooth cluster component and a few clumps
associated with individual galaxies.
The background source population is distributed as follows: they are allocated random
positions, number counts are generated in consonance with galaxy counts
typical for a 2 hour integration time in the R-band; 
the magnitudes are assigned by 
drawing the number count observed with the \textsc{cfht} and the mean redshift per magnitude bin
is derived from the \textsc{hdf} prescription; 
the shapes are assigned by drawing 
the ellipticity from a Gaussian distribution similar to the observed \textsc{cfht}
ellipticity  distribution. Typical density is 17 per arcmin$^{2}$.
We then ray-trace the lensing configuration using the publicly available \textsc{lenstool} 
\footnote{http://www.oamp.fr/cosmology/lenstool} software
to get a catalogue of images that we analyse as describe in next subsection.

\subsection{Methodology}
The details of the method have been given in Paper I.  
Here we just give a brief outline of the method.
Two methods are used in galaxy-galaxy lensing analysis. The most direct method
consists of obtaining an average shear field by simply binning up the
shear in radial bins from the centre of the lens outwards, and stacking
many individual galaxy shear profiles to
obtain a signal and to constrain an average galaxy halo population. Such
a method is possible when studying isolated field galaxies, since it makes the
assumption that we are able to isolate a lens in order to study it. However, this
assumptions is a very unlikely one, and galaxy-galaxy lensing is fundamentally
a multiple deflection problem, even when studying field galaxies. 
This was first pointed out by \textsc{bbs} in their
early work, who found that more than 50\% of their source galaxies should
have been lensed by two or more foreground galaxies: the closest lens on the sky
to any given source was not necessarily the only lens, and neither the
strongest lens. It has become clear
that it is almost never a unique lens that is responsible for the
detected lensing signal and that there are indeed no clean lines of
sight. Consequently, the problem is best tackled using an 'inverse'
method, and analysing galaxy-galaxy lensing using maximum likelihood
techniques is an example of such a method that we briefly expose below.\\
Once we have an image catalogue, we process it through a numerical code
that retrieves the input parameters of the lenses using a maximum
likelihood method as proposed by Schneider \& Rix, 1998 \cite{rix}.  For each image ($i$),
given a mass model for the foreground lenses galaxies (\emph{e.g.}
$\sigma_0$, $r$), we can evaluate the amplification matrix $a_i$ as a
contribution of all the foreground galaxies $j$ ; $z_j<z_i$ that lies
within a circle of inner radius $R_{\mathrm{min}}$, and outer radius
$R_{\mathrm{max}}$ and of centre the position of the image ($i$):
\begin{equation}
a_i(\sigma_0,r)=\sum_{\begin{array}{c} {z_j<z_i} \\ d(i,j)<R_{\mathrm{max}}\end{array}} a_{ij}(\sigma_0,r)
\end{equation}
Given the observed ellipticity $\vec{\varepsilon^i_{\mathrm{obs}}}$ and the associated amplification
matrix $a_i$, we are able to retrieve the intrinsic ellipticity $\vec{\varepsilon_i^s}$
of the source before lensing:
$
\vec{\varepsilon_i^s}=F\left(\vec{\varepsilon^i_{\mathrm{obs}}}, a_i(\sigma_0,r)\right)
=\vec{\varepsilon_i^s}(\sigma_0,r).
$
In order to assign a likelihood to the parameters used to describe the lenses
galaxies, we use
$P^s$, the ellipticity probability distribution in the absence of
lensing (a Gaussian of width 0.2).  Doing that for each image of the catalogue, we construct the
likelihood function:
$
\mathcal{L}(\sigma_0,r)=\prod_i P^s(\vec{\varepsilon^s_i})
$
which is a function of the parameters used to define the mass model
of the lenses. For each pair of parameters, we can compute a
likelihood.
The larger this function, the more likely the parameters used to describe
the lenses.  
\subsection{Results}

We present the results obtained for the simulated data set, for the
\textsc{piemd} and \textsc{nfw} models in Fig.~2. The point marks the value 
of the \textsc{input} parameters
used in order to generate the simulated catalogue, and the cross
stands for the value of the \textsc{output} parameters as estimated from the
maximum likelihood analysis.
Contours represent the 3$\sigma$, 4$\sigma$, 5$\sigma$ confidence levels, and along the
dotted lines, the mass within a projected radius $R_{\mathrm{aper}}\,=\,100$ kpc is
constant, equal to the value indicated on the plot.
The conclusion of this theoretical analysis is that we are able to retrieve the
\textsc{input} parameters
used in order to generate the simulated catalogue, and that the aperture mass
is constrained accurately.

\begin{figure}[h!] 
\begin{center}
\includegraphics[width=7cm,height=7cm]{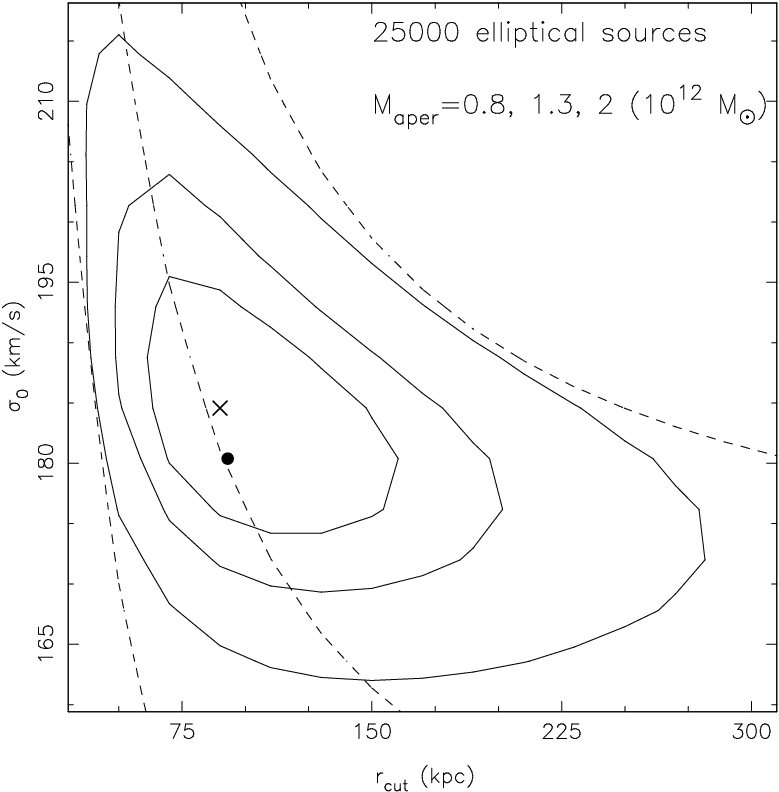}
\includegraphics[width=7cm,height=7cm]{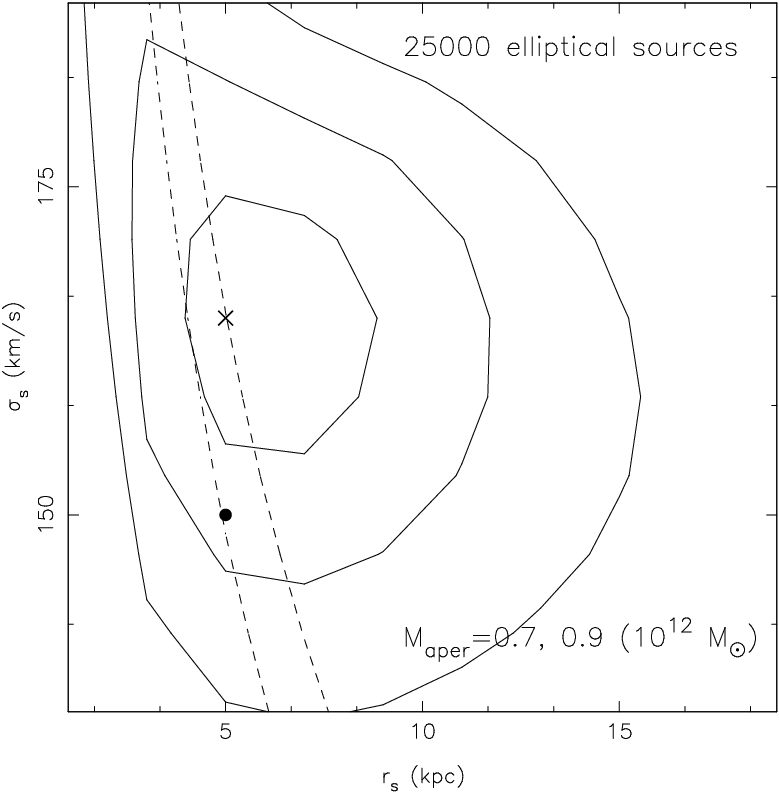}
\end{center}
\label{simures}
\caption{Results of the maximum likelihood analysis on simulated data set: left, using
a \textsc{piemd} profile; right, using a \textsc{nfw} profile.
The point marks the value of the \textsc{input} parameters
used in order to generate the simulated catalogue, and the cross
stands for the value of the \textsc{output} parameters as estimated from the
maximum likelihood analysis}
\end{figure}

\section{Application to a sample of cluster lenses at $z\sim0.2$}
Given an extensive simulation work described in Paper I, we were confident to apply this
maximum likelihood analysis to a sample of massive cluster lenses at $z\sim0.2$, with
the aim to constrain the properties of dark matter halos of galaxies living in such
a high density environment.
\subsection{Data and cataloguing}
The data were taken at the \textsc{cfht} with the \textsc{cfh12k} camera through
the B, R and I filters.
A detailed description of the data acquisition and reduction can be found in
Czoske, 2002 \cite{olliphd}. 
The object detection is described in Bardeau et~al., 2005 \cite{bardeau05}. 
Basic procedures have been implemented, using \textsc{sextractor}
\cite{Bertin-Arnouts1996} for object detection and magnitude computations, and
the \textsc{im2shape} software
developed by Bridle et~al., 2001 \cite{bridle01} for \textsc{psf} substraction and shape 
parameters measurements.  A master catalogue is produced which matches the objects
detected in the three filters and which contains colour indices built
from aperture magnitudes. This catalogue is used to plot the colour-magnitude
diagrams from which the sequence of elliptical is identified and
extracted. These elliptical cluster galaxies are the lenses for the background population.
We used only objects detected in all three bands and with reliable shape information, and
we undertook a photometric study to derive a redshift estimation for each background galaxy.

\subsection{Bayesian photometric redshift}
We used the \textsc{hyperz} \cite{hyperz} software to derive photometric redshifts.
Getting photometric redshifts with three bands is quite challenging
but possible and reliable for certain redshift ranges which are well
constrained by the filters we have.
Adding a prior probability allows to get better constraints than we would have without any
assumptions. The method implemented here has been developed by Benitez et~al, 1999 \cite{benitez}. 
The idea is to add a prior probability
which is not used by \textsc{hyperz} and in which we are confident.
We will use as a prior the luminosity function of a given
galaxy.  The final redshift assigned to a galaxy is determined by
combining the information coming from the \textsc{hyperz} probability
distribution with the prior probability distribution.
Basically, adding this prior allows us to get rid of some degeneracies in
the redshift probability distribution coming from \textsc{hyperz} due to using only
three filters.
From a theoretical analysis (see Paper II for details), we found that our filters
are well suited to constrain redshifts from $z\sim0.5$ to $z\sim1.5$, i.e. for the
background population. On the other hand, redshifts below 0.5 are not well constrained
by our filters.\\ 
To verify the reliability of the Bayesian photometric redshift estimation, we
compare with the \textsc{deep2} redshift survey (Coil et~al., 2004 \cite{deep2}):
they propose a simple colour-cut designed to select galaxies at $z>0.7$. As
discussed in Davis et~al., 2005 \cite{davis}, this colour-cut has proven effective.
%it results in a sample with $\sim$90\% of the objects at $z>0.7$, missing
%only $\sim$5\% of the $z>0.7$ galaxies.
We check to see where our objects with
$z_{\mathrm{bayes}}>0.7$ and $z_{\mathrm{bayes}}<0.7$ fall in a
colour-colour diagram, with respect to this colour-cut.  Figure
\ref{comparit} shows the results. The dashed line represents the
colour-cut: objects at $z>0.7$ are supposed to be above the regions
defined by these three dashed lines according to the \textsc{deep2} colour-cut.
The points represent our objects for which we have estimated
$z_{\mathrm{bayes}}<0.7$.  The crosses represent our objects for which
we have estimated $z_{\mathrm{bayes}}>0.7$. We can see from this plot
that our estimation of the redshift agrees well with the
colour-cut.

\subsection{Results}
\begin{figure}[h!] 
\begin{center}
\includegraphics[width=7cm,height=7cm]{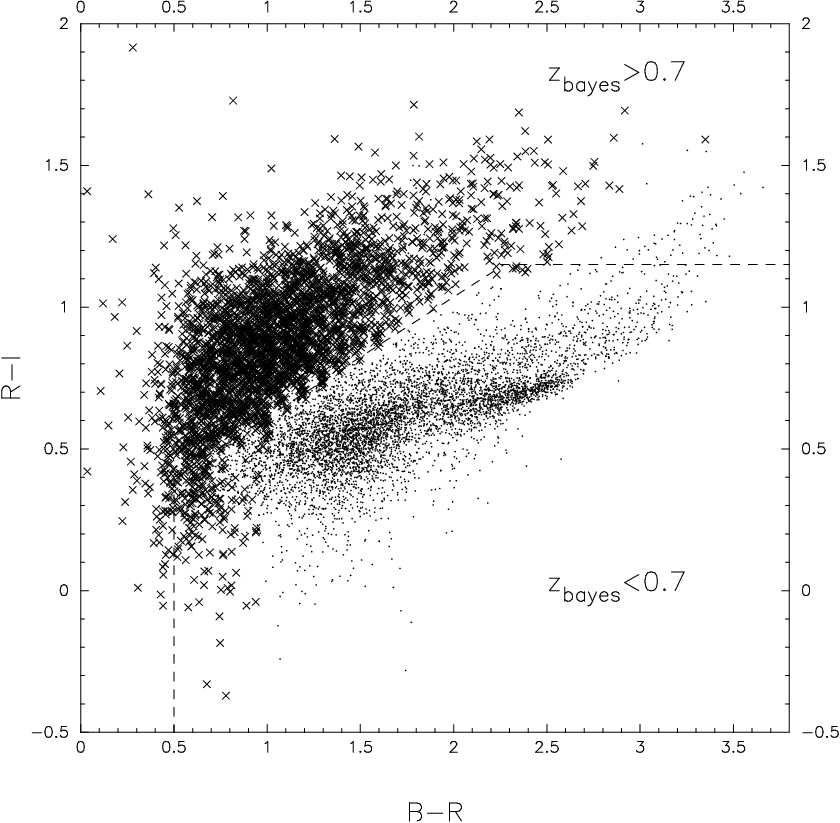}
\includegraphics[width=7cm,height=7cm]{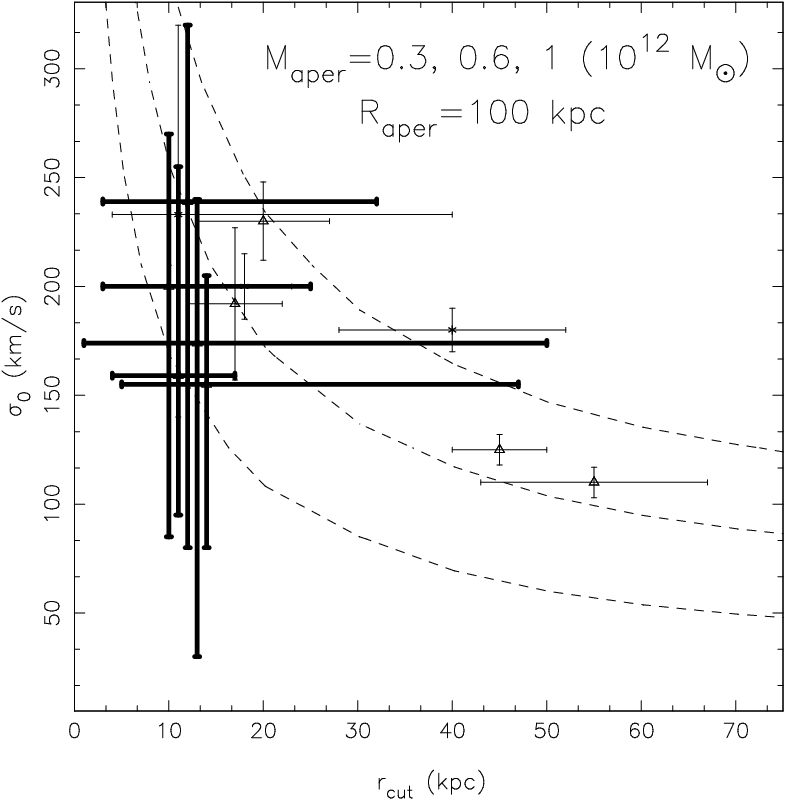}
\end{center}
\caption{
Left: Objects of our Abell~1763 catalogue with $z_{\mathrm{bayes}}>0.7$ (crosses)
and with $z_{\mathrm{bayes}}<0.7$ (dots).  The line
represents the simple colour-cut used in the \textsc{deep2} survey to select
objects with $z>0.7$ and $z<0.7$.
Right: comparison of our results (black) with the results
from Natarajan et al. and Geiger \& Schneider (grey)}
\label{comparit}
\end{figure}

We applied the maximum likelihood method to the image catalogues.
Lenses where extracted from the colour magnitude diagram, and objects with
redshift spanning from $z\sim0.5$ to $z\sim1.5$ where used as the background population.
The smooth component corresponding to the cluster itself was also considered; we put a large
clump with some parameters derived from a weak lensing analysis by Bardeau et al., 2006 \cite{bardeau06}.
Table~\ref{table1} summarises the
results we obtained using the two different profiles to fit the
deformations, and Fig.~\ref{comparit} shows our results (in black) with other results
on cluster galaxies (in grey), in the ($\sigma_0,r_{\mathrm{cut}}$) plane.\\
The main results are the
following: (i) we fit reasonable values for the velocity dispersion,
around 200 km\,s$^{-1}$. This is reasonable in the sense that it is
comparable to values inferred using more traditional methods (rotation
curves, X ray observations, satellite dynamics) (ii) we find dark
matter halos to be very compact compared to field galaxies of
equivalent luminosity: considering all cluster galaxies halos, an
upper limit on the truncation radius is set at 50 kpc (\textsc{piemd} results
on Abell~383), when the truncation radius inferred on field galaxies is
found to be larger than a few hundreds of kpc (see
Fig.~\ref{comparfieldcluster}). The truncation radius is related to the extension
of the halo. We can say that cluster galaxies are more compact than field
galaxies because the slope of their mass profile steepens earlier, thus the
corresponding mass profile reaches a low density value earlier.
As a consequence, dark matter halo of cluster galaxies are found to be less extended than
they are in the case of field galaxies.
\noindent The detection with an \textsc{nfw} profile
also points out a small spatial extent for the halos: the characteristic
radius $r_s$ is found to be smaller than a few kpc, which corresponds to a concentration
parameter $c>20$ in agreement with numerical simulations from Bullock et~al., 2001 \cite{bullock}.

\begin{table*}
\centering{
\begin{tabular}[h!]{ccccc}
\hline
\hline
\noalign{\smallskip}
Cluster & $\sigma_0^*$, {\footnotesize km\,s$^{-1}$ (\textsc{piemd})} & $r_{\mathrm{cut}}^*$, {\footnotesize kpc (\textsc{piemd})} & $r_\mathrm{s}^*$, {\footnotesize kpc (\textsc{nfw})} & {\footnotesize (M/L)$^*$} \\
\noalign{\smallskip}
\hline
\hline
\noalign{\smallskip}
\noalign{\smallskip}
A1763 & 200$^{+70}_{-115}$ (3$\sigma$) & $\leq$ 25 (3$\sigma$) & $\leq$ 2.5 (3$\sigma$) & 19$^{+16}_{-6}$ (3$\sigma$) \\
\noalign{\smallskip}
\noalign{\smallskip}
\hline
\noalign{\smallskip}
\noalign{\smallskip}
A1835 & 240 $^{+81}_{-159}$ (2$\sigma$) & $\leq$ 32 (2$\sigma$) & $\leq$ 5 (2$\sigma$) &  20$^{+18}_{-6}$ (3$\sigma$) \\
\noalign{\smallskip}
\noalign{\smallskip}
\hline
\noalign{\smallskip}
\noalign{\smallskip}
A2218 & 200$^{+96}_{-64}$ (1$\sigma$)& $\leq$ 18 (1$\sigma$) & $\leq$ 1 (1$\sigma$) &  13$^{+10}_{-12}$ (2$\sigma$) \\
\noalign{\smallskip}
\noalign{\smallskip}
\hline
\noalign{\smallskip}
\noalign{\smallskip}
A383 & 175 $^{+66}_{-143}$ (2$\sigma$) & $\leq$ 50 (2$\sigma$) & $\leq$ 2.2 (2$\sigma$) &  20$^{+13}_{-10}$ (3$\sigma$) \\
\noalign{\smallskip}
\noalign{\smallskip}
\hline
\noalign{\smallskip}
\noalign{\smallskip}
A2390 & 155$^{+50}_{-75}$ (1$\sigma$) & $\leq$ 47 (1$\sigma$) & $\leq$ 7 (1$\sigma$) &  10$^{+21}_{-4}$ (1$\sigma$)\\
\noalign{\smallskip}
\noalign{\smallskip}
\hline
\hline
\end{tabular}
\caption{Summary of the detections, for a $L^*$ luminosity. The mass corresponds to the total mass
computed with a PIEMD profile. Here $\sigma$ corresponds to the confidence level of the detection}
\label{table1}
}
\end{table*}

\section{Conclusion}

We have given a revue of galaxy-galaxy lensing detections obtained so far, both
on field galaxies as well as on cluster galaxies.
We have presented a maximum-likelihood analysis of galaxy-galaxy lensing
on simulated data to study the accuracy with
which input parameters for mass distributions for galaxies can be extracted..
We have applied this method on a sample of massive galaxy clusters and we
have derived some constraints on the dark matter halos of the 
elliptical cluster galaxies.
It is the first time that cluster galaxies are successfully probed from ground based
observations.
The main result of this work is to find galaxy halos in clusters to be significantly
less massive and more compact compared to galaxy halos around field galaxies of equivalent
luminosity.
This is in good agreement with previous galaxy-galaxy lensing studies. Moreover, this
confirmation is based on the analysis of 5 massive clusters lenses
whose properties are close one to each other, hence the confirmation
we provide is a strong one since it relies on a homogeneous sample.\\
This observational result is in good agreement with numerical simulations,
in particular with the tidal stripping scenario.
The theoretical expectation is that the global tidal field of a
massive, dense cluster potential well should be strong enough to
truncate the dark matter halos of galaxies that traverse the cluster
core (Avila-Reese et~al., 2005 \cite{avila05}; Ghigna et~al., 2000 \cite{ghigna00}; 
Bullock et~al., 2001 \cite{bullock}).

\section*{Acknowledgments}
Dark Cosmology Centre is funded by the Danish National Research
Foundation. PN acknowledges support from NASA via HST grant HST-GO-09722.06-A.
ML acknowledges Laurence Tresse and Sophie Maurogordato 
for organising this meeting.

\end{document}